\documentclass[ reprint,  amsmath,amssymb,  aps, ]{revtex4-1}%
\usepackage{graphicx}
\usepackage{dcolumn}
\usepackage{bm}
\usepackage{amsmath}
\usepackage{amsfonts}
\usepackage{amssymb}%
\providecommand{\U}[1]{\protect\rule{.1in}{.1in}}

\begin{document}


\title{Perpendicular  magnetic anisotropy at Fe/Au(111) interface studied by M\"{o}ssbauer, x-ray absorption, and photoemission spectroscopies }%

\author{Jun Okabayashi$^{1}$, Songtian Li$^{2}$, Seiji Sakai$^{2}$, Yasuhiro Kobayashi$^{3}$, Takaya Mitsui$^{4}$, 
Kiyohisa Tanaka$^{5}$, Yoshio Miura$^{6}$, and Seiji Mitani$^{6,7}$}%

\affiliation{
$^1$Research Center for Spectrochemistry, The University of Tokyo, Bunkyo-ku, Tokyo 113-0033, Japan \\
$^2$National Institutes for Quantum and Radiological Science and Technology QST, Takasaki, Gunma 370-1292, Japan \\
$^3$Institute for Integrated Radiation and Nuclear Science, Kyoto University, Kumatori, Sennan-gun, Osaka 590-0494, Japan \\
$^4$National Institutes for Quantum and Radiological Science and Technology QST, Sayo-cho, Hyogo 679-5148, Japan \\
$^5$UVSOR, Institute of Molecular Science, Okazaki, Aichi 444-8585, Japan \\
$^6$Research Center for Magnetic and Spintronic Materials, National Institute for Materials Science (NIMS), Tsukuba 305-0047, Japan \\
$^7$Graduate School of Science and Technology, University of Tsukuba, Tsukuba 305-8577, Japan
}

\date{\today}%
%

\begin{abstract}%

The origin of the interfacial perpendicular magnetic anisotropy (PMA) induced in the ultrathin Fe layer on the Au(111) surface was examined using synchrotron-radiation-based M\"{o}ssbauer spectroscopy (MS), X-ray magnetic circular dichroism (XMCD), and angle-resolved photoemission spectroscopy (ARPES). To probe the detailed interfacial electronic structure of orbital hybridization between the Fe 3$d$ and Au 6$p$ bands, we detected the interfacial proximity effect, which modulates the valence-band electronic structure of Fe, resulting in PMA. MS and XMCD measurements were used to detect the interfacial magnetic structure and anisotropy in orbital magnetic moments, respectively. $In$-$situ$ ARPES also confirms the initial growth of Fe on large spin-orbit coupled surface Shockley states under Au(111) modulated electronic states in the vicinity of the Fermi level. This suggests that PMA in the Fe/Au(111) interface originates from the cooperation effects among the spin, orbital magnetic moments in Fe, and large spin-orbit coupling in Au. These findings pave the way to develop interfacial PMA using $p$-$d$ hybridization with a large spin-orbit interaction. 

\end{abstract}%

\maketitle

\section{\label{sec:level1}Introduction}

One of the fascinating aspects of material design research is interfacial atomic control, which creates a variety of novel properties. Recently, materials designed using artificial-intelligent machine learning processes have also been anticipated to provide a clue to the combination of elements and atomic orientation possessing novel properties \cite{1,2,3}, which has garnered tremendous attention in the spintronics research and high-performance device applications. In particular, when ferromagnetic transition metals (TMs) are deposited on surfaces with large spin-orbit coupling or topologically insulating properties, novel properties such as perpendicular magnetic anisotropy (PMA) emerge at the interfaces, which are derived from the Rashba-type spin-orbit coupling effects through interfacial symmetry breaking \cite{4,5,6,7}. The interfacial magnetic proximity can be a novel technology for developing and enhancing spin-orbit-related functionalities. Therefore, hybrid structures combining different spin-orbit coupling strengths yield a magnetic cooperative effect as enhancements of the interfacial orbital magnetic moments, resulting in a large PMA. The spin-orbit coupling between the ferromagnetic 3$d$ TMs Fe or Co and the more-than-half 4$d$-5$d$ heavy TM elements of non-ferromagnetic materials such as Pd and Pt have been utilized for PMA through the proximity at the interfaces \cite{8,9,10}. In the case of Au, the 5$d$ states are mostly  occupied. The relationship between the Rashba-type spin-orbit interaction in Au(111), mainly originating from the 6$p$ Shockley surface states, and PMA in 3$d$ TMs has not been explicitly clarified. Here, we address the Fe/Au(111) interfaces using a magnetic Fe layer and non-magnetic Au because the proximity between a large Rashba-type spin-orbit coupling constant on the Au surface and the spins in the Fe layer results in interfacial PMA. Historically, the interfaces of Fe/Au(111) have been thoroughly examined for the development of multilayered structures with PMA \cite{11,12,13,14}. However, there are few reports on microscopic and element-specific investigations of PMA in Fe/Au(111) interfaces.

Gold Au (111) surfaces have been extensively investigated using scanning tunneling microscopy \cite{15,16,17,18} and angle-resolved photoemission spectroscopy (ARPES) \cite{19,20} because this type of surface exhibits not only free-electron-like parabolic $k^2$ Shockley surface states, but also a large Rashba-type spin-orbit coupling originating from ${\nabla}{\phi}{\times}{\hbar}k$ with a linear term of crystal momentum ${\hbar}k$ and a gradient of potential profile $\phi$. The large spin-orbit interaction in the heavy element gold provides a wide variety of topological physics, and spin-orbit coupled phenomena at the surfaces and interfaces through Rashba-type splitting \cite{6}.

To investigate the layer-resolved electronic and magnetic states in Fe films with PMA, M\"{o}ssbauer spectroscopy (MS) using a $^{57}$Fe isotope source is a powerful and unique tool because it detects the information of doped $^{57}$Fe sites at the interface. Because the nuclear excitation interacts with the electrons of the $s$ orbital in the Fe atom, the electronic structure of the Fe 3$d$ states can be probed through the $s$-$d$ coupling. Although there have been several previous reports of MS in Fe thin films possessing in-plane magnetic anisotropy, they are mainly limited to detection by conversion electron MS using isotope sources \cite{21,22,23,24}. Recent developments in MS using synchrotron radiation (SR) incident beams have opened up a unique technique using focused and polarized beams, and precise measurements, which have enabled us to separately determine the contributions from the interface and other bulk regions \cite{25}. Although SR-based MS can be employed for a layer-resolved discussion, there have been few studies on interfacial PMA.

ARPES using SR beams is a powerful technique for detecting detailed electronic structures at the surfaces and interfaces within the probing depth. Numerous reports have investigated the electronic structures of deposited ultrathin layers on Rashba-type Au(111) surface states \cite{26,27,28}. X-ray absorption spectroscopy (XAS) and X-ray magnetic circular dichroism (XMCD) can directly deduce the spin and orbital magnetic moments based on their angular dependence and yield studies of element-specific magnetic anisotropy. Although the initial growth mechanism for depositing less than one monolayer (ML) has been examined \cite{29}, studies for ultrathin Fe layers in a few ML range, exhibiting PMA, are necessary owing to their potential application in spintronic devices.

In this study, we adopted MS, XMCD, and ARPES to investigate Au/Fe interfaces to elucidate the layer-resolved electronic and magnetic properties. Using SR-based MS, we aim to depict the interfacial electronic states. Using ARPES with a first-principles calculation, we aim to understand the electronic structures at the magnetic interfaces on the Rashba-type Au (111) surface in order to advance research on novel PMA using spin-orbit coupled interfaces.

\section{\label{sec:level1}Experimental}

All sample preparations were conducted using electron-beam evaporation. A 100-nm thick Au layer was grown on the $c$-plane sapphire substrate. The surface of a (111)-oriented Au crystal was cleaned by argon ion bombardment and subsequent annealing at 400 $^{\circ}$C. This procedure is known to form well-ordered surfaces. Reflection high-energy electron diffraction (RHEED) and low-energy electron diffraction (LEED) were employed to probe the surface flatness. 
The Fe layer was grown at 150 $^{\circ}$C on a clean Au surface using  $^{56}$Fe and enriched $^{57}$Fe sources. The Au layer was deposited onto the ultrathin Fe layer at room temperature for MS, XMCD, and superconducting quantum interference device (SQUID) measurements.  The thicknesses of the individual layers were monitored using a quartz thickness monitor. MS was conducted at SPring-8 BL11XU, where a diamond X-ray phase plate and an iron borate nuclear Bragg monochromator were installed to generate polarized beams. The beamline details are described in Ref. \cite{30}. 
Focused beams of 14.4 keV for nuclear excitation of $^{57}$Fe were irradiated at a grazing angle of 0.16$^{\circ}$ with respect to the surface plane of the sample in several polarization modes under a magnetic field applied along the surface normal. A cooling system using liquid He was used at a temperature of 20 K. The M\"{o}ssbauer spectra were calibrated with reference to an $\alpha$-Fe foil. 

XMCD measurements were conducted using BL-7A at the Photon Factory of the High-Energy Accelerator Organization (KEK). A magnetic field of $\pm$1 T was applied along the incident polarized beam by switching the magnetic field directions. The absorption signals are defined as ${\mu}^+$ and ${\mu}^-$. XAS and XMCD are defined as (${\mu}^+$+${\mu}^-$)/2 and ${\mu}^+-{\mu}^-$, respectively. The total electron yield mode was adopted. The XAS and XMCD spectra were obtained after normalization with respect to the incident photon intensities \cite{31}. All XAS and XMCD measurements were performed at 80 K using liquid N$_2$.

ARPES measurements were performed at the beamline BL5U of the UVSOR synchrotron facility at the Institute for Molecular Science, Okazaki, Japan. The BL5U system was equipped with a Monk-Gillieson VLS-PGM monochromator and a hemispherical electron energy analyzer possessing two-dimensional image detection with an acceptance angle of ${\pm}$15$^{\circ}$ for measurements with $p$-polarized (horizontal) and $s$-polarized (vertical) SR beams. Linearly polarized SR beams were injected at 45$^{\circ}$ from the sample surface normal. 
The ARPES end station is connected to the sample preparation chamber to transfer the deposited sample without breaking an ultra-high vacuum. The surface conditions were monitored by LEED and core-level photoemission intensities of Fe 3$p$ and Au 4$f$ with a photon energy of 120 eV. Photoemission spectra were calibrated using the Fermi edge of gold in angle-integrated spectra, and the energy resolution was set to 15 meV. 

\section{\label{sec:level1}Results and Discussion}

We synthesized the stacked structures of Au (1 nm; capping)/ Fe (0.5 nm; 3 ML) on an Au(111) substrate. Figure 1(a) shows the RHEED patterns for each deposition stage. Mainly, 1$\times$1 patterns were observed, which originated from the Au(111) substrate. The Fe deposition on Au(111) proceeds with the crystalline growth keeping the fcc Fe(111) structure up to 3 ML. 
For an Fe layer thicker than the critical thickness, the growth of bcc Fe with the (110) orientation proceeds \cite{18}. In a previous report, for the initial growth stage, Fe island growth was observed using scanning tunneling microscopy \cite{18} and analyzed by grazing incidence x-ray diffraction \cite{32}. With increasing thickness, coalescence occurs, and the layer-by-layer continuous film growth starts accompanied by the interfacial strain \cite{29}. The magnetization measurements by SQUID exhibit a clear ferromagnetic behavior for the 3-ML-thick Fe with capping Au layer. Further, as shown in Fig 1(b), clear PMA was detected in agreement with a previous report of ultrathin Fe on the Au(111) surface \cite{13}. The magnetic anisotropy energy ($E_\mathrm{MA}$), which is determined from the product surrounded by in-plane and out-of-plane magnetic hysteresis ($M$-$H$) curves, is estimated to be $1.3{\times}10^5$ J/m$^3$, including the shape magnetic anisotropy contribution.

Figure 2 shows the M\"{o}ssbauer spectra of 3-ML-thick Fe, where the layer consists of a 1-ML-thick enriched $^{57}$Fe facing the Au and 2-ML-thick $^{56}$Fe layer before Au capping. Magnetic ordering can be monitored by hyperfine sextet-line shapes ($I_i$; $i=1-6$) through the transition from the nuclear spin quantum number 1/2 to 3/2 states. As shown in Fig. 2(a), in the case of $\pi$-polarized beam injection, where the magnetic field vector of the incident linearly polarized SR beam  ($\bf{B}$) is aligned in-plane to the film, as shown in the inset, the 2nd and 5th peak intensities ($I_2$ and $I_5$) are suppressed under an external magnetic field of 0.03 T applied in the perpendicular direction. Because $\bf{B}$ does not couple with the spins in the in-plane direction ($y$), the sample exhibits PMA or is parallel to the beam direction ($z$). In the case of a non-polarized beam, as shown in Fig. 2(b), $I_2$ and $I_5$ intensities appear, and the line shapes become similar to an isotropic bulk case. From Figs. 2(a) and 2(b), the $z$ direction is determined to be the magnetic easy axis. Furthermore, as shown in Fig. 2(c), without applying a magnetic field, $I_2$ and $I_5$ almost disappear with small intensities, which probes the remanent magnetization in the $M$-$H$ curves. This suggests that the sample of the 3-ML-thick Fe layer on Au(111) exhibits PMA. 

Next, to elucidate the detailed magnetic characteristics, we conducted spectral fitting of the M\"{o}ssbauer spectra. Because of the broadening of the $I_1$ and $I_6$ peaks, two types of components were adopted for spectral fitting. 
The M\"{o}ssbauer parameters, isomer shift (IS), quadrupole splitting (QS), and hyperfine magnetic field ($H_\mathrm{hf}$) derived from the fitting for each spectrum in Fig. 2 are listed in Table I. 
The component with a larger $H_\mathrm{hf}$ (Component 1) is assigned as the interfacial Fe-Au component, whereas the other corresponds to the Fe bulk (Component 2). These two components had similar area intensities. To analyze the PMA from the spectra, the intensity ratio of $I_{2,5}$ ($I_2$ or $I_5$) and $I_{3,4}$ is related to the tilted angle $\theta$ from the sample surface normal, as follows: 
\begin{equation}
{\frac{I_{2,5}}{I_{3,4}}}={\frac{4\mathrm{sin}^2{\theta}}{1+\mathrm{cos}^2{\theta}}}.
\end{equation}
From the intensity ratio in Fig. 2, the angle $\theta$ is close to zero, suggesting a perpendicular magnetic easy axis, which is consistent with the $M$-$H$ curves in Fig. 1(b). We emphasize that the detection of PMA at the Fe/Au interface can be achieved using SR-based MS with detailed information of the interfacial Fe ions. 
Furthermore, the interfacial charge transfer effect due to the difference in the electronegativity between Au and Fe can be elucidated from the IS values, which are estimated to be 0.23$\pm$0.02 mm/s, for the interfacial component referring to high-symmetric $\alpha$-Fe. 
Quadrupole splitting is also modulated at the interface regions because of the large electric-field gradient from Au,  determined mainly by $I_{3,4}$ splitting. Two types of QS components are related to hyperfine splitting. These values differ from the bulk values. The large QS values originate from the interfacial electric field gradient due to the Fe-Au bonding, and the other is from the contribution of the Fe-Fe bonding. 

Figure 3 shows the XAS and XMCD of Fe $L$-edges with angular dependence taken at 80 K. Clear metallic line shapes are detected even at a 3-ML thickness. In the normal incident case, the spin ($m_\mathrm{s}^{\perp}$) and orbital magnetic moments ($m_\mathrm{orb}^{\perp}$) along the normal direction are deduced in the relation $m_\mathrm{orb}^{\theta}=m_\mathrm{orb}^{\perp}+(m_\mathrm{orb}^{\parallel}-m_\mathrm{orb}^{\perp})\mathrm{sin}^2\theta$. Because of the $d$ orbital states, the angular dependence is expressed as a function of $\mathrm{sin}^2{\theta}$ \cite{33,34}. The oblique incident geometry case includes half of the in-plane component for $m_\mathrm{orb}^{\parallel}$. By adopting the XMCD sum rules, $m_\mathrm{orb}^{\perp}$ and $m_\mathrm{orb}^{\parallel}$ are determined to be 0.10 and 0.09 ${\mu}_\mathrm{B}$, respectively, with error bars of $\pm$10\% due to the ambiguity in the subtraction of the background. The value of $m_\mathrm{orb}^{\perp}$ is enhanced through the interfacial proximity with Au. The value of $m_s$ is expressed using the magnetic dipole moment ($m_\mathrm{T}$) as $m_\mathrm{s}^{\mathrm{eff} {\theta}}=(m_\mathrm{s}-7m_\mathrm{T}^{\perp})+\frac{21}{2} m_\mathrm{T}^{\perp} \mathrm{sin}^2{\theta}$, where $m_\mathrm{T}^{\perp}=0$ in the case of a magic angle $\mathrm{cos}^2{\theta}=1/3$. The value of $m_s$ is 1.71 ${\mu}_B$ and a negligible value of $m_\mathrm{T}$ is deduced using the  Fe 3$d$ hole number of 6.9 estimated from the density-functional-theory (DFT) calculation. These values are consistent with the previous XMCD of Fe/Au for more than 2-ML-thick Fe on Au. For less than 2 ML, it has been reported that the contribution of $m_\mathrm{T}$ is enhanced \cite{28}. The element-specific magnetic anisotropy energy $E_\mathrm{MA}$ from the orbital moment anisotropy and spin density anisotropy is estimated from the following relation within the second order perturbation for spin-orbit coupling:  
\begin{equation}
E_{\mathrm{MA}} {\sim} \frac{1}{4}\xi \left( m_\mathrm{orb}^{\perp}-m_\mathrm{orb}^{\parallel} \right) -\frac{{\xi}^2}{{\Delta}E_\mathrm{ex}} \left[ \frac{21}{2}\frac{3}{2} m_\mathrm{T}^{\perp} +\alpha \right]
\end{equation}
using ${\xi}_\mathrm{Fe}$= 50 meV for Fe and exchange splitting ${\Delta}E_\mathrm{ex}$, resulting in 125 $\mu$eV/atom, which is translated into the order of $10^4$ J/m$^3$ without using the second term corresponding to $m_\mathrm{T}^{\perp}$. The positive values of $E_\mathrm{MA}$ stabilize the PMA. The value of $\alpha$ is the residual after being expressed by $m_\mathrm{T}$, which is negligible. The relation of $m_\mathrm{T}^{\perp}+2m_\mathrm{T}^{\parallel}=0$ is assumed. The value of $E_\mathrm{MA}$ from XMCD is overestimated compared with that from SQUID in general by a pre-factor of 0.1 \cite{33}. Considering the pre-factor, our results suggest that the orbital moment anisotropy of Fe through proximity to Au is essential for the enhancement of $m_\mathrm{orb}^{\perp}$. Further, the $M$-$H$ curves at the Fe $L_3$-edge are also shown in Fig. 3 (b). The PMA was clearly observed. We note that the element-specific $M$-$H$ curves in XMCD can exclude the diamagnetic contribution from the substrates, which is difficult to correct for SQUID data, especially in the case of ultra-thin films. Therefore, the orbital moment anisotropy in the Fe sites is dominant for the PMA in the Fe/Au interface because of its proximity to the Au layer. 

ARPES reveals the valence-band electronic structure of an ultrathin Fe layer under the Au (111) Shockley surface states. 
As shown in Fig. 4(a), the angle-integrated valence-band photoemission spectra with the Fe layer dependence account for the changes in the spectral line shapes, suggesting the appearance of Fe 3$d$ states. In the corresponding band dispersion mapping, the Au (111) Shockley surface states exhibited parabolic bands  and whole valence-band dispersions were observed at a photon energy of 45 eV. On this surface, Fe deposition gradually modulates the surface states, and the Fe 3$d$ states appear with increasing Fe layers in the flat band feature owing to orbital degeneracy. At the initial growth stage, the valence-band electronic structures of Au are modulated by Fe deposition and cannot be explained by simple summations of the Au and Fe line shapes. 
As shown in Fig. 4(b), the expanded view in the vicinity of the Fermi level ($E_\mathrm{F}$) in Au clearly shows the Rashba-type spin-orbit splitting in the parabolic surface state, which is the same as the previous report \cite{26,27,28}. 
In the case of horizontal incident beam mainly detecting $x^2-y^2$, $3z^2$, and $zx$ orbitals, as shown in Fig. 4(c), a high-intensity region appears around the $\Gamma$ point taken at a photon energy of 60 eV, where the perpendicular component of the wave vector $k_{\perp}$ crosses near the $\Gamma$ point, assuming the unit cell in the band-structure calculation, and the Fe 3$d$ cross-section is enhanced. The parallel component $k_{\parallel}$ of 1.0 $\mathrm{\AA}^{-1}$ corresponds to the midpoint between the $\overline{\Gamma}$-$\overline{K}$ line. Quite small dispersive features were observed even in the 3-ML-thick Fe deposition. In contrast, in the case of a vertical incident beam detecting $xy$ and $yz$ orbitals, non-dispersive broad bands appear. 
The band dispersions estimated from the DFT calculation are also plotted. 
Because of the initial growth stage, distinct band dispersion features are not formed at a thickness of 3 ML. The DFT calculations agree with the broad ARPES data only at a qualitative level. The discrepancy in the dispersions between the ARPES spectra and the DFT calculation might be due to the effect of surface disorder at the initial growth stage before forming the band structure of bulk Fe \cite{35}. 

A first-principles band-structure calculation for 3-ML-thick Fe on Au(111) was carried out by adjusting the lattice constant to Au. At the initial growth stage of Fe on Au(111), this assumption is plausible through the proximity in the ultrathin layer case before the lattice relaxation to stabilize the bulk bcc structure. DFT calculations were performed using the Vienna $ab$ $initio$ simulation package \cite{36} with the projector augmented wave potential \cite{37}, including spin-orbit interaction and the spin-polarized generalized gradient approximation for the exchange and correlation term \cite{38}. The Fe/Au(111) slab was constructed with 19 atomic layers of Au and three atomic layers of Fe with an in-plane lattice constant of 4.078/$\sqrt{2}$ (${\mathrm{\AA}}$) of the hexagonal unit cell. A plane-wave cut-off energy of 400 eV for the wave function and $35{\times}35{\times}1$ $k$-points in the first Brillouin zone. Figure 5 shows the orbital-resolved band dispersions of the Fe 3$d$ states with Au valence bands along the $\overline{\Gamma}$-$\overline{K}$ symmetric line in the two-dimensional hexagonal Brillouin zone. The Au(111) band dispersions with Rashba-type surface state splitting were clearly observed. The dispersive features derived from Fe are typical for dispersions in strained fcc structures. 
The 3$d_{z^2}$ orbitals at the $\overline{\Gamma}$ point are evident for the  $p$-$d$ hybridization along the out-of-plane direction. In the vicinity of $E_\mathrm{F}$, the in-plane orbitals are dominant, which favors the out-of-plane orbital magnetic moment. 
In the case of bcc Fe, the band dispersion features were quite different \cite{39}. 
Therefore, the Fe layer facing on the Au is essential for the PMA from the viewpoint of DFT calculations.

Considering the above results, we discuss the detailed electronic and magnetic properties at the Fe/Au interface in the following five subjects. 

$First$, the hyperfine field in the MS is reconsidered, as listed in Table I. It consists of the Fermi contact term ($H_\mathrm{Fermi}$), dipole interaction ($H_\mathrm{dipole}$), and orbital magnetic moment ($H_\mathrm{orb}$) through the electron orbital. 
\begin{equation}
H_\mathrm{hf}=H_\mathrm{Fermi} + H_\mathrm{dipole} + H_\mathrm{orb}.
\end{equation}
The term $H_\mathrm{Fermi}$ is described by the $\delta$-function at the nuclear center and is dominant for $H_\mathrm{hf}$ compared with the terms of $H_\mathrm{dipole}$ and $H_\mathrm{orb}$ at the $^{57}$Fe site \cite{40}. As the two components have almost equal intensities, the $^{57}$Fe layer faces the interface with Fe-Fe and Fe-Au chemical bonds.  The enhancement of $H_\mathrm{hf}$ at the interfacial component is explained by Fe-Au bonding because the Au solute in the Fe$_{1-x}$Au$_{x}$ alloy influences to increase $H_{\mathrm{hf}}$ \cite{41}.  The difference in $H_\mathrm{hf}$ of 3 T corresponds to 0.17 meV energy. The $H_\mathrm{orb}$ term also contributes to the anisotropy in orbital magnetic moments through the 4$s$-3$d$ hybridization at the atomic $^{57}$Fe site. 

$Second$, the quadrupole term corresponding to the QS at the interface is discussed. As shown in Fig. 6, the interface modulates the potential profile, which induces an electric field of $-{\nabla}{\phi}$, which is related to the effective magnetic field of $-{\nabla}{\phi}{\times}{\hbar}k$ and enhances the Rashba-type spin-orbit interaction ${\sigma}[{\nabla}{\phi}{\times}{\hbar}k]$ using a spin matrix $\sigma$. The interfacial Fe site is affected by the large potential gradient, resulting in a large quadrupole splitting in the MS measurements, as listed in Table I. The large quadrupole splitting in MS is specific for the Fe/Au interface compared with that of the Fe/MgO interface \cite{21}. 

$Third$, the interfacial band structures were compared with first-principles calculations. According to the second-order perturbation theory, the PMA is stabilized at the Fe/Au interfaces for the (100) and (111) orientations \cite{42,43}. Because the Au 5$d$ states are almost completely occupied, the Au 6$p$ and Fe 3$d$ states are dominant in the vicinity of $E_\mathrm{F}$. The in-plane 3$d$ orbitals crossing the $E_\mathrm{F}$ mainly contribute to the PMA, which is similar to the Fe/MgO interface recognized as a $p$-$d$ hybridized interface \cite{44}. In the case of the Fe/Au interface, not only $p$-$d$ hybridization but also Rashba-type spin-orbit coupling affects the PMA. 


$Finally$, for the origin of the PMA in the Fe/Au interfaces, the Au surface states of the $p$ bands stabilize the PMA of the Fe 3$d$ spins. Therefore, the Fe/Au interface is categorized as a $p$-$d$ interface with a Rashba-type spin-orbit interaction using the interfacial potential gradient. This is different from the $d$-$d$ interfaces of the Fe/Pt and Co/Pd cases because the 5$d$ (4$d$) states are not significant origins for the PMA in the Fe/Au interface.   \\

\section{\label{sec:level1}Summary}

We investigated the microscopic origin of interfacial PMA induced in an ultrathin Fe layer on a Au(111) surface using synchrotron-radiation-based MS, XMCD spectroscopy, and ARPES. We detected the interfacial proximity effect, which modulates the valence-band electronic structure of Fe, resulting in PMA. The polarization dependence in MS under a magnetic field shows the PMA characteristics and the enhancement of the interfacial component in the hyperfine field through Fe-Au bonding. $In$-$situ$ ARPES and DFT calculations also confirm that the initial growth of Fe on the strongly spin-orbit coupled Shockley surface states in Au(111) modulates the electronic states in the vicinity of the $E_\mathrm{F}$. This suggests that the PMA in the Fe/Au(111) interface originates from the cooperation between spin, orbital magnetic moments in Fe, and Rashba-type spin-orbit coupling in Au. 

\begin{acknowledgments}
This work was partially supported by JSPS KAKENHI (Grant No. 16H06332), the foundation of the Toyota Physical and Chemical Institute, and the Shimazu Science and Technology Foundation. Parts of the synchrotron radiation experiments were performed with the approval of the QST Advanced Study Laboratory, SPring-8 (proposal  Nos.  2019A3551, B3551 and 2020A3551), the Photon Factory Program Advisory Committee, KEK (No. 2019G028), and UVSOR proposal Nos. 19-556 and 20-758.
\end{acknowledgments}

\begin{table*}[htb]
\begin{center}
\caption{Parameters derived from the fitting of M\"{o}ssbauer spectra displayed in Fig. 2 (a) - (c). Two types of components are assumed. Parameters of the isomer shift (IS), quadrupole splitting (QS), and hyperfine field ($H_\mathrm{hf}$) are listed. 
}
\begin{tabular}{c|c|c|c|c|c|c} \hline
 & \multicolumn{3}{c|}{Component 1} & \multicolumn{3}{c}{Component 2}  \\ \hline 
 & IS (mm/s) & QS (mm/s) & $H_\mathrm{hf}$ (T) & IS (mm/s) & QS (mm/s) & $H_\mathrm{hf}$ (T) \\ \hline 
(a) $\pi$-polarization (0.03 T) & 0.23(6) & -0.13(12) & \hspace{0.5cm} 36.9(1) \hspace{0.5cm} & 0.29(5) & -0.08(10) & \hspace{0.5cm} 33.4 \hspace{0.5cm} \\ \hline 
(b) non-polarization (0.03 T) & 0.20(2) & -0.10(3) & 36.5(2) & 0.31(2) & -0.02(3) & 33.3(2) \\ \hline 
(c) $\pi$-polarization (0 T) & 0.25(5) & -0.03(8) & 36.1(2) & 0.33(4) & 0.11(8) & 33.4(2) \\ \hline 
\end{tabular}
\end{center}
\end{table*}

\begin{figure}
[ptb]
\begin{center}
\includegraphics[
width=4in
]%
{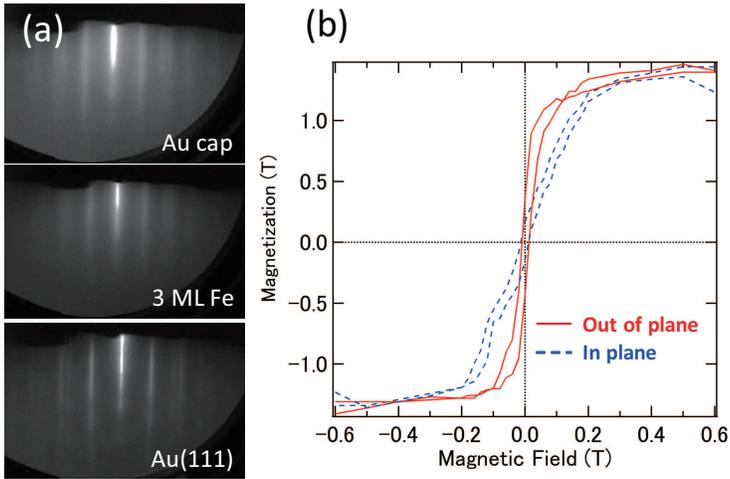}%
\end{center}
\caption{(Color online) Characterization of sample structures: (a) RHEED patterns after cleaning the Au substrate, 3-ML thick Fe deposition, and 1 nm-thick Au capping layer deposition. (b) Magnetization curves applying the magnetic fields along the out-of-plane and in-plane directions.}
\end{figure}

\begin{figure}
[ptb]
\begin{center}
\includegraphics[
width=5in
]%
{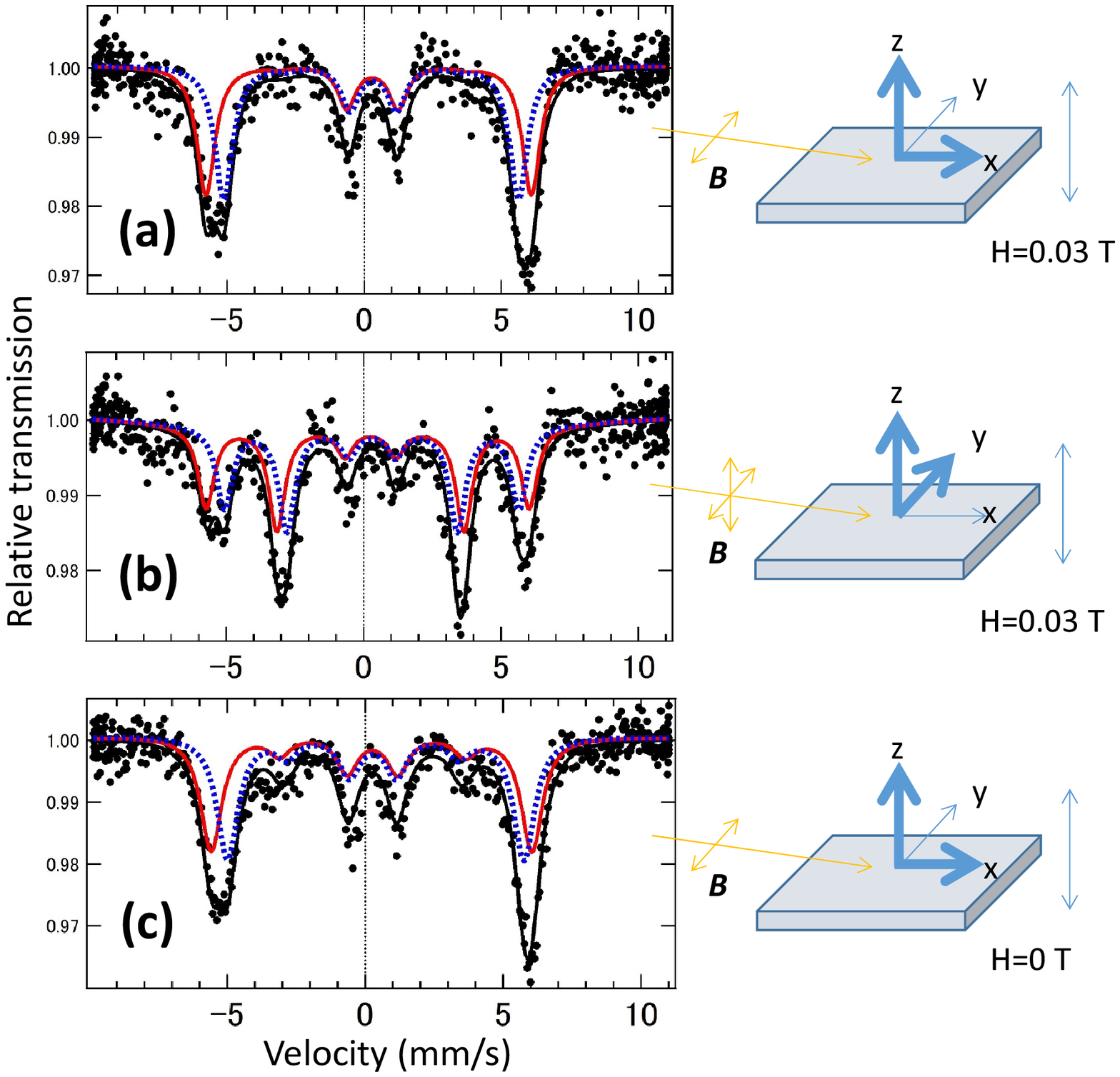}%
\end{center}
\caption{(Color online) Synchrotron-radiation M\"{o}ssbauer spectra of Au (1 nm)/Fe (3 ML)/ Au(111) structure detected under application of a magnetic field of 0.03 T along the perpendicular direction to the sample surface normal using a (a) $\pi$-polarized beam and (b) non-polarized beam at 20 K. (c) The case without applying a magnetic field, detected using a remanent state. Solid and dot curves are the fitting results using two types of curves in red (Component 1) and blue (Component 2) color, respectively. }
\end{figure}

\begin{figure}
[ptb]
\begin{center}
\includegraphics[
width=5in
]%
{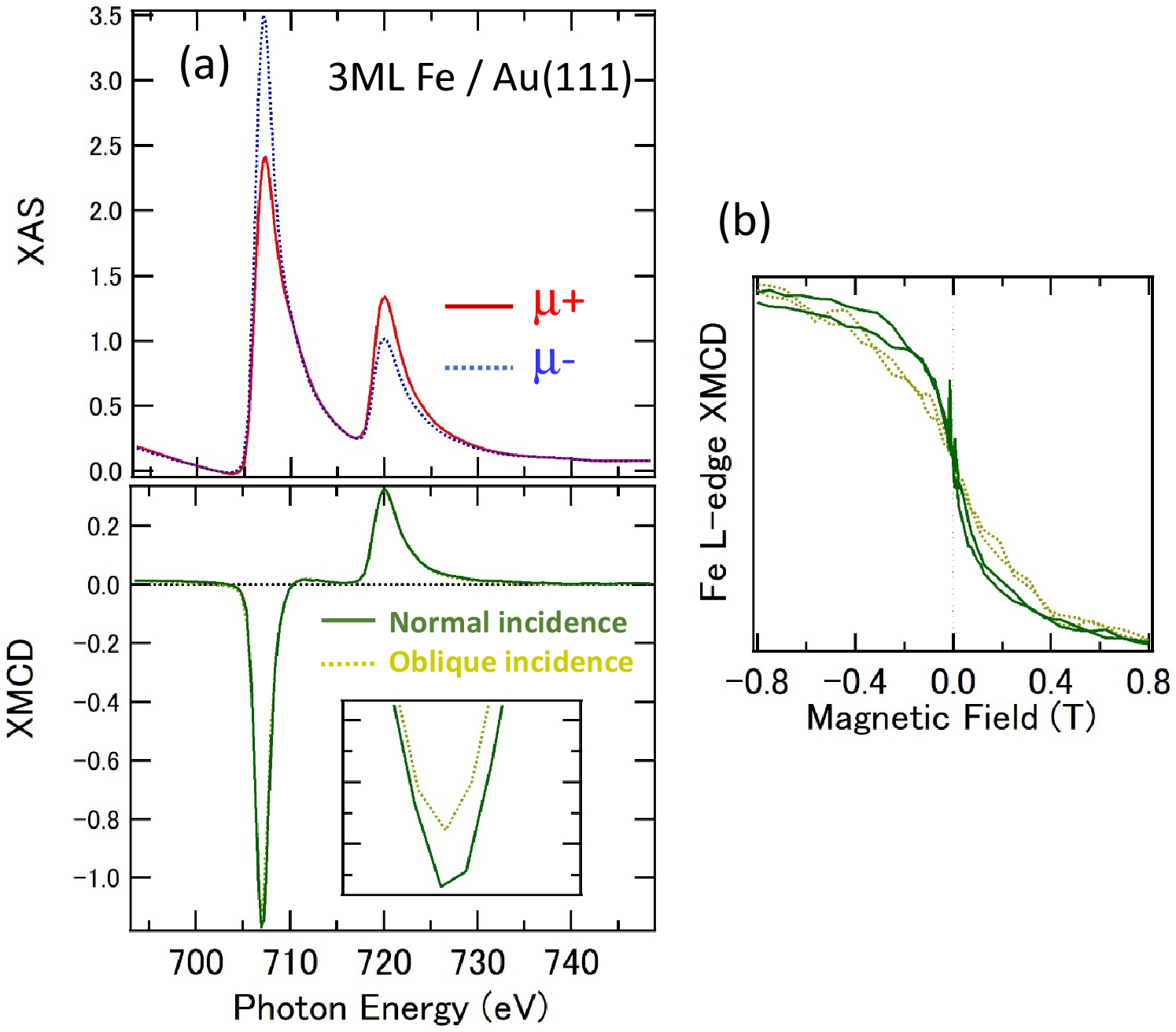}%
\end{center}
\caption{(Color online) XAS and XMCD of Fe $L$-edges detected by circularly polarized beams, ${\mu}^+$ and ${\mu}^-$: (a) XAS taken at normal incident geometry and XMCD by normal and oblique incident geometry tilted at 60$^\circ$. Inset shows the expanded view around $L_3$-edge. (b) Magnetic field dependence of XMCD at Fe $L_3$-edge photon energy in normal and oblique incident cases.}
\end{figure}

\begin{figure}
[ptb]
\begin{center}
\includegraphics[
width=5in
]%
{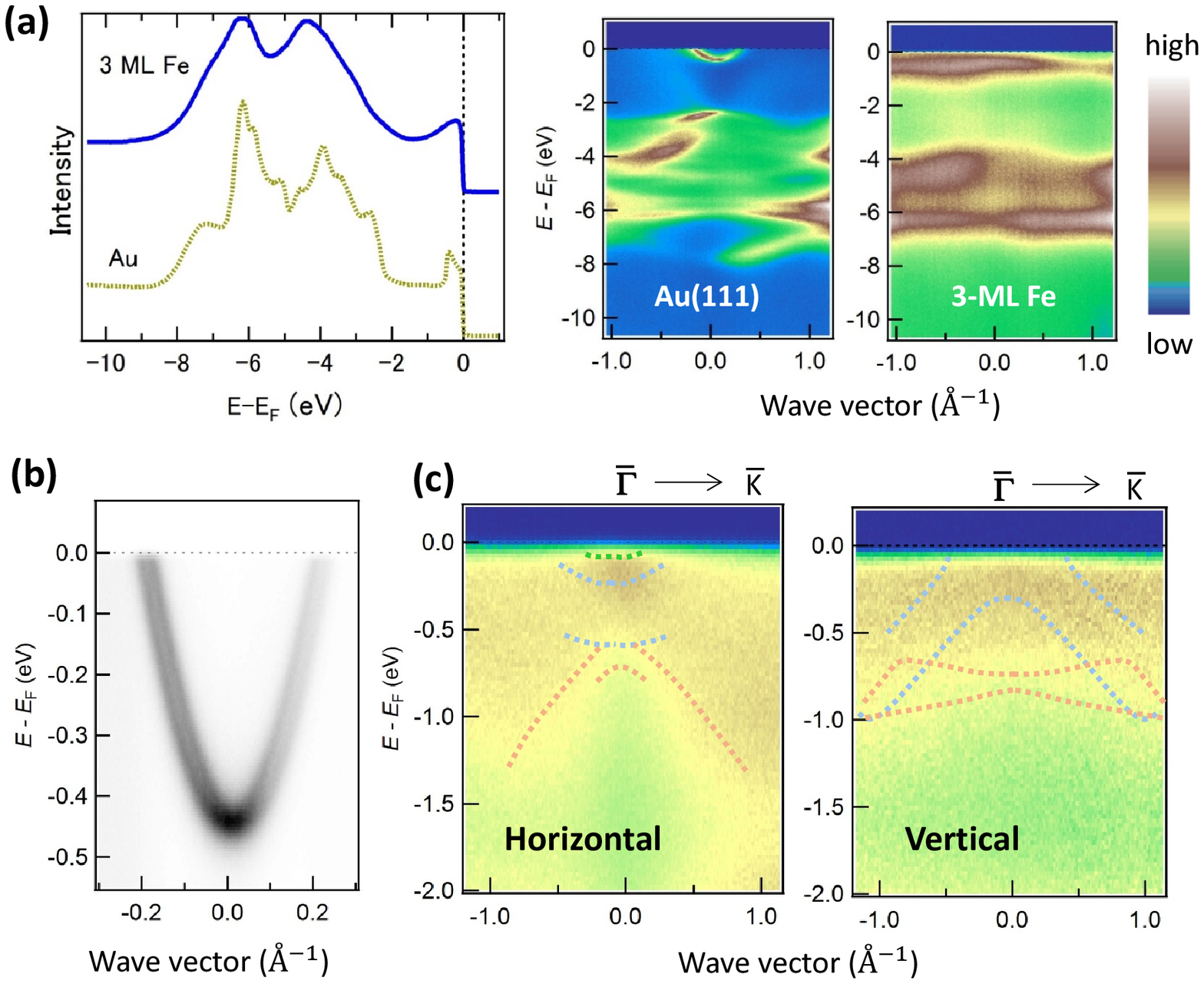}%
\end{center}
\newpage
\caption{Valence-band photoemission spectra of Au(111) and 3-ML-thick Fe on Au(111): (a) Angle-integrated valence band spectra referring to the Fermi level ($E_\mathrm{F}$) and corresponding band dispersion mapping in ARPES taken at a photon energy of 45 eV. (b) the mapping of angle-resolved spectra of Au (111) surface taken at 45 eV and (c) 3-ML-thick Fe in horizontal and vertical incident beams taken at a photon energy of 60 eV along the $\overline{\Gamma}-\overline{K}$ direction in the hexagonal surface Brillouin zone. Dot curves in (c) are the theoretical calculation results depicted from Fig. 5 with orbital characters of $d_{3z^2}$ (green), $d_{yz}$ and $d_{zx}$(red), and $d_{x^2-y^2}$ and $d_{xy}$ (blue).  }   
\end{figure}

 \begin{figure}
[ptb]
\begin{center}
\includegraphics[
width=5in
]%
{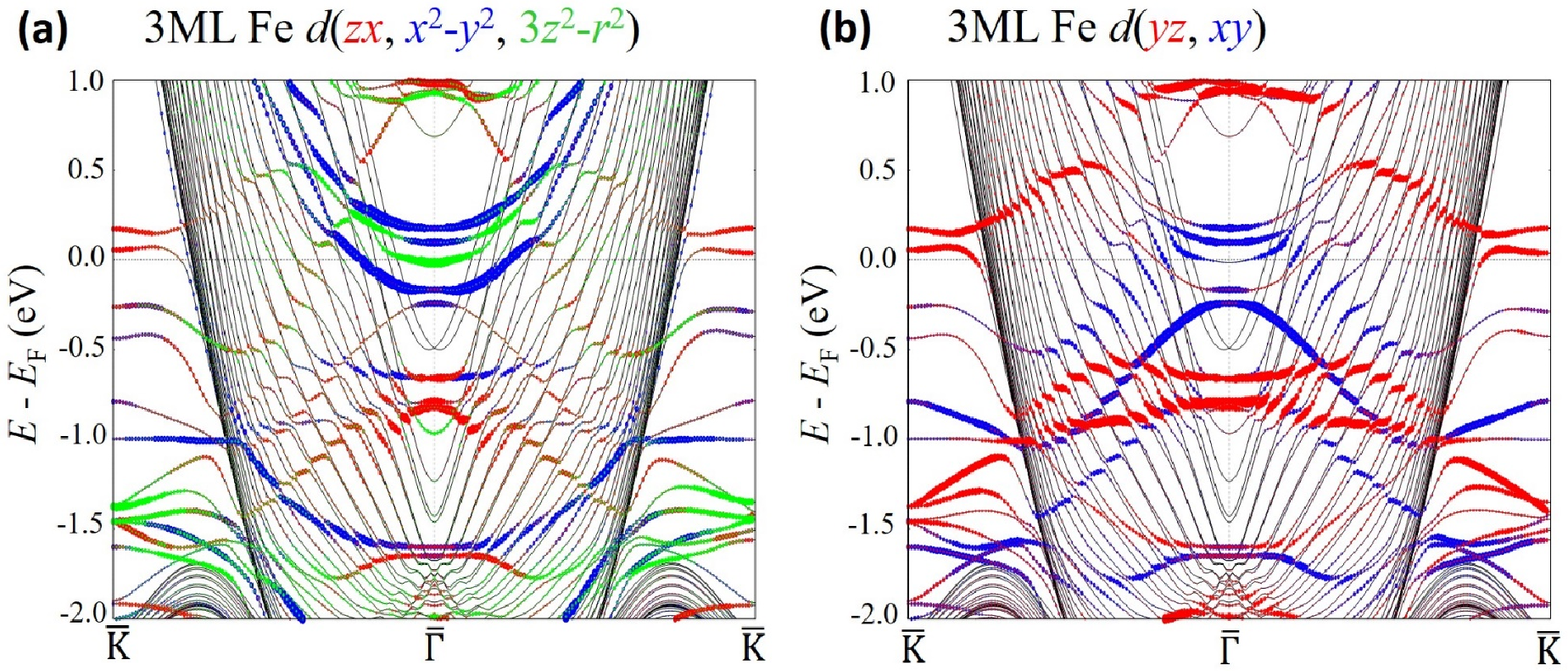}%
\end{center}
\caption{A first-principles calculation of orbital-resolved band dispersions of 3-ML-thick Fe on Au(111). The highlighted bands in color are the Fe 3$d$ states. Black curves correspond to the band dispersions of Au. (a) $d_{x^2-y^2}$ (blue), $d_{3z^2}$ (green), and $d_{zx}$ (red) orbitals are highlighted as the horizontal geometry in ARPES. (b) $d_{xy}$ (blue) and $d_{yz}$ (red) orbitals are highlighted corresponding to the vertical geometry in ARPES.   }   
\end{figure}

\begin{figure}
[ptb]
\begin{center}
\includegraphics[
height=8in,
width=6in
]%
{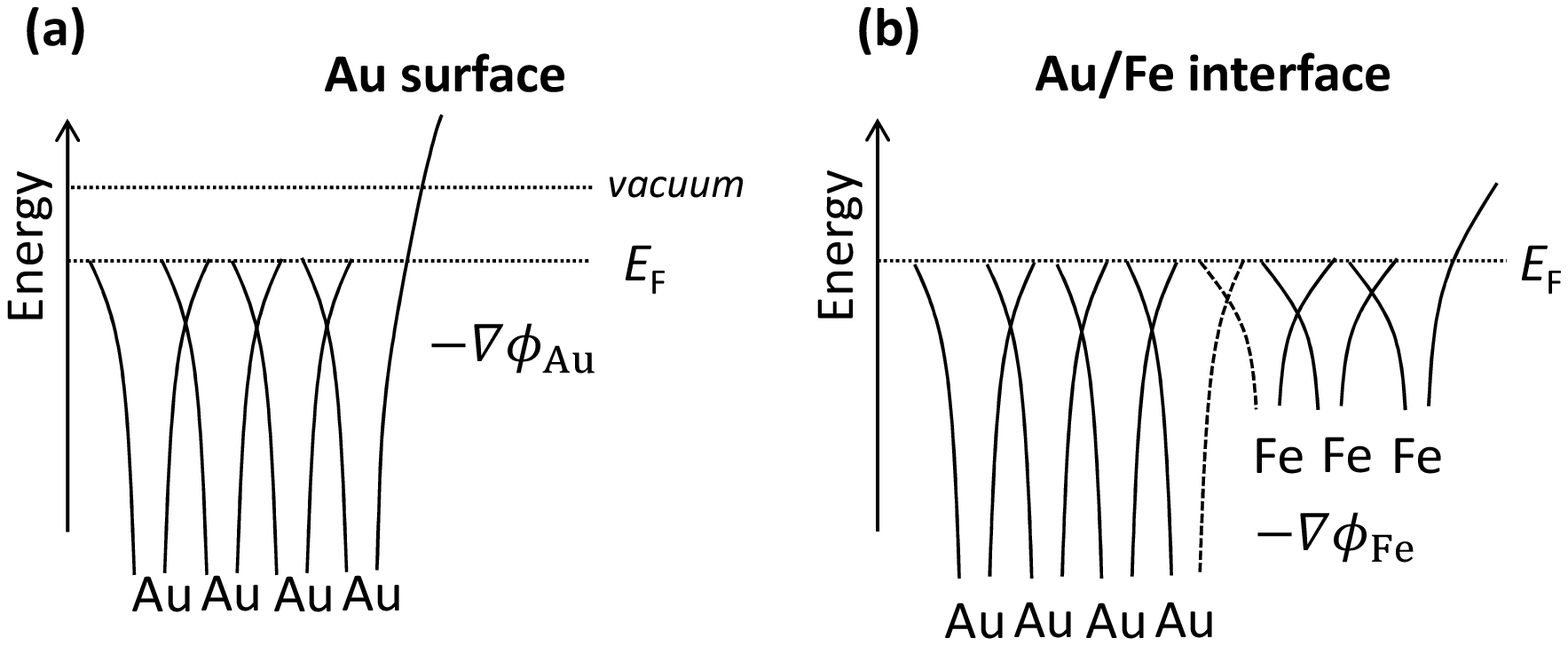}%
\end{center}
\caption{Schematic illustrations of potential profile. (a) Potential profile of Au surface and (b) that of Fe/Au interface.}
\end{figure}

\end{document}